\documentclass[aps,prl,superscriptaddress,twocolumn,tightenlines]{revtex4}

\usepackage {graphicx}
\usepackage {amsmath}
\usepackage {amsfonts}
\usepackage {amssymb}
\usepackage {mathrsfs}
\usepackage {bm}

\usepackage {epstopdf}

\newcommand {\be}{\begin{eqnarray}}
\newcommand {\ee}{\end{eqnarray}}
\newcommand {\rmd} {{\rm d}}  
\newcommand{\bk}{{\bf k}}

\newcommand{\bq}{{\bf q}}
\newcommand{\br}{{\bf r}}

\newcommand{\thalf}{\tfrac{1}{2}}

\begin{document}

\title {Valley density-wave and multiband superconductivity in Fe-pnictides}
\author {Vladimir Cvetkovic}
\affiliation {Department of Physics \& Astronomy, The Johns Hopkins University,
Baltimore, MD 21218}
\author {Zlatko Tesanovic}
\affiliation {Department of Physics \& Astronomy, The Johns Hopkins University,
Baltimore, MD 21218}

\date {\today}

\begin{abstract}
The key feature of the Fe-based superconductors is
their quasi 2D  multiband Fermi surface.
By relating the problem to a negative $U$ Hubbard model
and its superconducting ground state, we show
that the defining instability of such a Fermi surface
is the valley density-wave (VDW), a {\em combined} spin/charge density-wave
at the wavevector connecting the electron and hole valleys.
As the valley parameters change by doping or pressure,
the fictitious superconductor  experiences ``Zeeman splitting'',
eventually going into a non-uniform 
``Fulde-Ferrell-Larkin-Ovchinikov'' (FFLO) state,
an {\em itinerant} and often {\em incommensurate} VDW of the real world,
characterized by the metallic conductivity from the ungapped remnants of the Fermi surface. 
When ``Zeeman splitting'' exceeds
the ``Chandrasekhar-Clogston'' limit, the ``FFLO'' state disappears,
and the VDW is destabilized. 
Near this point, 
the VDW fluctuations and interband pair repulsion are essential ingredients of
high-$T_c$ superconductivity in Fe-pnictides.

\end{abstract}


\maketitle

\section{Introduction}

Recently, the superconductivity below 7 K
in LaOFeP \cite {LaOFeP} led to the discovery of high  $T_c \sim$ 26 K in its
doped sibling LaO$_{1-x}$F$_{x}$FeAs ($x > 0.1$) \cite{LaOFeAs}.
Even higher $T_c$'s were found by replacing 
La with other rare-earths (RE), up to
the current record of $T_c =$ 55 K \cite {recordTc}. 
These are the first
non-cuprate superconductors exhibiting such high $T_c$'s and
their discovery has touched off a storm of activity
\cite{PhysicaC}.

In this paper, we introduce a new element into the theoretical debate
by considering a unified model of spin density-wave, orbital density-wave,
 structural deformation 
and superconductivity in Fe-pnictides.
The model is simple but it contains the necessary physical features.
The essential ingredient are electron and hole pockets
(valleys) of the quasi two-dimensional (2D)
multiply-connected Fermi surface (FS) \cite{Lebegue,Mazin,Hirschfeld}. 
To extract the basic physics
we consider spinless electrons first,
and only a single electron and a single hole
band with identical band parameters. We then 
show that this model can be related
to a 2D {\em negative} $U$ Hubbard model, the ground state of which
is known exactly -- it is a superconductor \cite {negativeU}. In real FeAs materials, this 
fictitious superconductivity translates
into a fully gapped valley density-wave (VDW), a unified state representing
a combination
of spin, charge and orbital density-waves (SDW/CDW/ODW) at
the {\em commensurate} wavevector ${\bf M}$ connecting the two valleys.
Next, we introduce two different ``chemical potentials'', 
$\mu^e\not = \mu^h$ for the electron
and the hole valleys -- this describes the effect of doping the parent
iron-pnictide compounds and corresponds to the external
Zeeman splitting in our fictitious negative $U$ Hubbard model. As
$\delta\mu= \mu ^e - \mu^h$ increases, so does this Zeeman splitting,
and eventually our fictitious superconducting state approaches to and exceeds
the ``Chandrasekhar-Clogston'' limit, giving way
to a {\em non-uniform} Fulde-Ferrell-Larkin-Ovchinikov (FFLO) ground state
at an {\em incommensurate} wavevector ${\bf q}$, where
$|{\bf q}|$ is set by 
$\delta k_F=k_F^e-k_F^h$, and thus by doping $x$. This ``FFLO state''
is nothing but an incommensurate (IC) VDW at the wavevector ${\bf M} + {\bf q}$.
Finally, as $\delta k_F~(x)$ exceeds certain critical value $\delta k_c~(x_c)$,
the ``superconducting'' state is completely destroyed and so is 
the VDW in a true material. However, for $\delta k_F$ above but
near $\delta k_c$, we consider strong ``superconducting'' fluctuations and
find that these VDW fluctuations can induce
{\em real} superconductivity in Fe-pnictides (see Fig. \ref{fig1}).
In principle, one could avoid the mapping to the negative-U Hubbard model
and argue that the VDW instability in pnictides occurs for the same
reasons as the SDW instability found in, say, Cr \cite {RiceHalperin}. We find,
however, that our `fictitious superconductivity' description is more appropriate to pnictides
not only due to its illustrative purposes, but also because it allows us to
extend the analogy to the ``FFLO'' state and multiband ``SC'', i.e., VDW. Furthemore,
it provides a natural venue, by using 
the known near-rigourous results on two-dimensional superconductors
\cite{Trubowitz}, to highlight the crucial role played in real superconductivity
by the interband pair-scattering processes, as dicussed at length below.

The appearance of an SDW in pnictides along with a structural transition, which
we argue to be a signature CDW coupled to phonons, has been established
early on in the so-called 1111 family \cite {OakRidge}, and 
the universal presence of these orders in other
families of pnictides and related materials (122, 11, etc.)
has been confirmed by many authors \cite {WBao122, McQueen11}.
Both the magnetic and the structural order set in at nearly identical
temperatures, the experimental fact that has motivated us 
to model this problem as a one major, ``mother''
instability (VDW) driven by a large energy scale, which is then split into
several stages (say, CDW, followed by SDW, and eventually ODW order)
by interaction terms considerably smaller in magnitude. A main feature
of our fictitious FFLO state are the ungapped portions of the reconstructed Fermi 
surface(s) (FS's). These FS's
have been observed in pnictides directly \cite {Analytis} as well as
indirectly through their signatures in metallic resistivity \cite {PrFeAsO}
and recently detected incommensurability of the SDW order \cite {JCDavis}.

\section{An idealized model of a valley density-wave and an FFLO state
in a fictitious superconductor}

The band structure of Fe-pnictides
\cite{Lebegue,Mazin,Hirschfeld} can be
parametrized by the five-orbital tight-binding model \cite{Kuroki, Cvetkovic}. The
key feature is the {\em multiband} nature of the
FS, possessing both hole and electron sections. 
We work with the properly defined Fe-pnictide unit cell which contains
two of Fe and two pnictide (As or such) atoms per unit cell.
The basic physics is captured by the Hamiltonian
describing two hole ($c^\alpha$) and two electron ($d^\beta$) bands 
centered at the $\Gamma$ and $M$ points of the 
2D Brillouin zone (BZ), respectively,
\be
H &=& H_0 + H_{int}~, \label {H} \\
H_0 &=& \sum_{\bk,\sigma, \alpha} \epsilon^{(\alpha)}_{\bk} c_{\bk,\sigma}^{(\alpha) \dagger} c_{\bk,\sigma}^{(\alpha)}+
\sum_{\bk,\sigma, \beta} \epsilon^{(\beta)}_{\bk} d_{\bk,\sigma}^{(\beta) \dagger} d_{\bk,\sigma}^{(\beta)}~, \label {H0} \\
  H_{int} &=& \thalf\int \rmd^2\br \rmd^2\br' V(\br,\br')n (\br )n(\br')~,\label{Hint} 
\ee
where $\sigma ,\sigma'=\uparrow ,\downarrow$ and $\alpha,\beta$ 
are the spin and band labels, respectively,
$\epsilon_\bk ^{(\alpha,\beta)}$ is the hole (electron) dispersion
near the FS, $V(\br,\br')$ is the effective
interaction and 
$n(\br)=\sum_\sigma \psi^\dag_{\sigma}(\br)\psi_{\sigma}(\br)$, with
$\psi_\sigma(\br)=\sum_{\bk,\alpha} c_{\bk,\sigma}^{(\alpha)}
\varphi_{\bk}^{(\alpha)}(\br)+
\sum_{\bk, \beta} d_{\bk,\sigma}^{(\beta)}\phi_{\bk}^{(\beta)}(\br)$.
$\varphi_{\bk}^{(\alpha)}(\br)$ and $\phi_{\bk}^{(\beta)}(\br)$ are the Bloch
wavefunctions of hole (electron) bands. 

For simplicity, 
(\ref{Hint}) includes only the screened density-density
repulsion; its form
becomes more complex if we integrate out the bands away from
the Fermi level $E_F$, generating additional interactions in the spin and
interband (orbital) channels. Furthermore, we could equally well
start from the minimal tight-binding representation of Ref.\ \cite{Cvetkovic}
and introduce the interaction term in the Wannier representation as
\be
H_{int} = \frac{1}{2}U_d\sum_i n_{di}^2 -J_{\rm Hund} \sum_i {\bf S}_{di}^2
+ (\cdots)~,
\label{HintWannier}
\ee
where $n_{di}$ and ${\bf S}_{di}$ are the total particle number and spin in
Wannier d-orbitals of iron. $U_d$ describes the overall Hubbard-like repulsion
on iron sites while $J_{\rm Hund}$ signifies the intra d-orbital Hund coupling.
In addition, there are numerous intra d-orbital interactions, as well
as various similar terms for p-orbitals on pnictide sites, all contributing to
$(\cdots)$ \cite{Sawatzky}. However, all such interaction terms feed into the generic
classes of quartic vertices near the FS which are generated by 
$H_{int}$ (\ref{Hint}); only the precise
numerical values of various vertices are affected. In particular,
as long as the influence of $J_{\rm Hund}$ (and $(\cdots)$ terms) is relatively small 
and one is in the weak-to-intermediate coupling regime argued to be relavant
to pnictides \cite{Cvetkovic}, the overall numerical hierarchy of energy
scales defined by these various classes of vertices remains intact, as descussed below.
Finally, we further simplify the problem by exploiting the
fact that all electron (hole) bands have $E_F$ near their bottom (top)
and their Fermi wavevectors $k_F$'s are $\ll M$. 
This allows us to restrict our attention to the first BZ and
take continuum limit, with $V(\br,\br')\to V(\br -\br')$.

$H_{int}$ (\ref{Hint}) generates three classes of vertices:
i) the intraband ($c^\dag c c^\dag c$ and $d^\dag d d^\dag d$),
ii) the interband ($ c^\dag c d^\dag d$), and 
iii) the mixed ($d^\dag c c^\dag d + {\rm h.c.}$ 
and $c^\dag d c^\dag d + {\rm h.c.}$).
All arise from $H_{int}\to \thalf\sum_{\bq}\tilde V_{\bq} n_{\bq} n_{-\bq}$,
where
\be
n_{\bq} &=&
\sum_{\bk \sigma\alpha\alpha'} \zeta^{(\alpha \alpha')}_{\bk+\bq,\bk}
c_{\bk+\bq, \sigma}^{(\alpha) \dagger} c_{\bk, \sigma}^{(\alpha')} +
\sum_{\bk \sigma \beta \beta'} \zeta^{(\beta \beta')}_{\bk+\bq,\bk}
d_{\bk+\bq, \sigma}^{(\beta) \dagger} d_{\bk, \sigma}^{(\beta')}
+ \nonumber \\
&&\sum_{\bk \sigma \alpha \beta} \gamma^{(\alpha \beta)}_{\bk+\bq, \bk}
c_{\bk+\bq, \sigma}^{(\alpha) \dagger} d_{\bk, \sigma}^{(\beta)} + {\rm h.c.}~,
\label {density}
\ee
$\tilde V_{\bq}$ is the Fourier transform (FT) of $V(\br -\br')$, and
\be
  \zeta^{(\alpha \alpha')}_{\bk, \bk'} &=& \int \rmd^2 \br
  e^{i (\bk'-\bk)\cdot\br}\varphi_{\bk}^{(\alpha)*}(\br)\varphi_{\bk'}^{(\alpha')}(\br), \nonumber \\
  \zeta^{(\beta \beta')}_{\bk,\bk'} &=& \int \rmd^2 \br
  e^{i (\bk' - \bk)\cdot\br}\phi_{\bk}^{(\beta)*}(\br)\phi_{\bk'}^{(\beta')}(\br), \label {zetas} \\
  \gamma^{(\alpha \beta) }_{\bk,\ bk'} &=& \int \rmd^2 \br
  e^{i (\bk' - \bk)\cdot\br}\varphi_{\bk}^{(\alpha)*}(\br)\phi_{\bk'}^{(\beta) }(\br). \nonumber
\ee
The following should be kept in mind about these three classes of vortices:
first, all exhibit considerable variation as one moves around the FS. This is
the consequence of significant variations in the orbital content of various
bands in different portion of the BZ. Second, we find that, generically,
the intra and the interband vertices are comparable in magnitude
while the mixed ones are notably smaller. This remains true regardless of whether we use
the interaction (\ref{Hint}), (\ref{HintWannier}) or some other related form
as long as $J_{\rm Hund}$ is not dominating the physics and $(\cdots)$ (\ref{HintWannier})
are relatively small.

To illustrate the latter claim, we include the Hund's coupling to the interaction terms and
compare the vertices with and without it. For example, the hole intraband vertex
\be
  U^{(\alpha)}_{\bk, \bk', \bq} c_{\bk + \bq, \sigma}^{(\alpha) \dagger}
    c_{\bk' - \bq, \sigma'}^{(\alpha) \dagger} c_{\bk' \sigma'}^{(\alpha)} c_{\bk \sigma}^{(\alpha)}
\ee
acquires strength
\be
  U^{(\alpha)}_{\bk, \bk', \bq} &=& \left ( V_\bq + \tfrac 1 4 J_{\rm Hund} \right )
  \zeta_{\bk + \bq, \bk}^{(\alpha \alpha)} \zeta_{\bk' - \bq, \bk'}^{(\alpha \alpha)} + \nonumber \\
  && \thalf J_{\rm Hund} \zeta_{\bk + \bq, \bk'}^{(\alpha \alpha)} \zeta_{\bk' - \bq, \bk}^{(\alpha \alpha)},
\ee
and therefore the Hund's coupling effectively only adds up to the Coulomb potential
in this scattering channel. The
same is true for other intraband scattering processes, the second mixed term $G_2$,
and most importantly for the interband scattering vertices
\be
  W^{(\alpha \beta)}_{\bk, \bk', \bq} &=& \left ( V_\bq + \tfrac 1 4 J_{\rm Hund} \right )
  \zeta_{\bk + \bq, \bk}^{(\alpha \alpha)} \zeta_{\bk' - \bq, \bk'}^{(\beta \beta)} + \nonumber \\
  && \thalf J_{\rm Hund} \gamma_{\bk + \bq, \bk'}^{(\alpha \beta)} \gamma_{\bk, \bk' - \bq}^{(\alpha \beta)*},
\ee
where the second term is negligible for small $\bq$. The only interaction vertex that is
more affected by $J_{\rm Hund}$ than the others is the first mixed term
\be
  G^{(\alpha \beta)}_{1 / \bk, \bk', \bq} &=& \left ( V_\bq + \tfrac 1 4 J_{\rm Hund} \right )
  \gamma_{\bk + \bq, \bk}^{(\alpha \beta)} \gamma_{\bk', \bk' - \bq}^{(\alpha \beta) *} + \nonumber \\
  && \thalf J_{\rm Hund} \zeta_{\bk + \bq, \bk}^{(\alpha \alpha)} \zeta_{\bk' - \bq, \bk'}^{(\beta \beta)}, \label {G1}
\ee
due to the relative size of $\zeta$'s and $\gamma$'s. 
However, as long as the screened Coulomb
potential is the strongest interaction (i.e., $J_{\rm Hund} \lesssim V_\bq$ here) the
changes to vertices due to the other sources of scattering (such as those
in Eq.\ \eqref {HintWannier}) will be only quantitative in nature.

\begin{figure}[tbh]
\centering
\includegraphics[width=0.6\columnwidth]{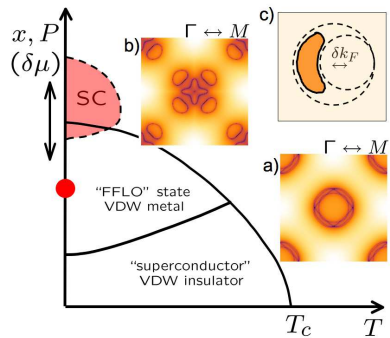}
\caption{\label{fig1} (Color online)
Phase diagram of Fe-pnictides, depicting the 
evolution of our fictitious superconductor from the
fully gapped VDW insulator to the ``FFLO superconductor'' -- a partially
gapped metallic VDW  -- to the {\em real} superconductor (SC) under the influence of 
the ``Zeeman splitting'' $\delta\mu$ (doping or pressure). Red dot symbolizes the 
parent compounds and the regime below it might be physically inaccessible.
Insets: FS of a) the normal state in the folded ($\Gamma\leftrightarrow M$) BZ \cite{Cvetkovic},
b) the VDW metal (computed with the interband interaction set to unity) -- this
is the $C_4$ version of c) the continuum FFLO state \cite{kulic}.
The remaining states are fully gapped.
}
\end{figure}

We now observe that the shapes of different sections
of the FS (Fig.\ \ref{fig1}) resemble each other to a reasonable degree.
Furthermore, various masses are also roughly similar \cite{Kuroki,Cvetkovic}. 
Thus, to make theoretical progress, it is useful to first
assume that all electron and hole bands are equal
$- \epsilon^{(\alpha = h1)}_{\bk} = - \epsilon^{(\alpha = h2)}_{\bk} =
\epsilon^{(\beta = e1)}_{\bk + {\bf M}} = \epsilon^{(\beta = e2)}_{\bk + {\bf M}} \equiv \epsilon_{\bk}^0$. 
After making the particle-hole (p-h) transformation
$d_{\bk,\sigma}^{(\alpha)}\to e_{\bk,\sigma}^{(\alpha)}$,
$c_{\bk,\sigma}^{(\alpha)}\to \sigma h_{\bk,-\sigma}^{(\alpha)\dagger}$,
the Hamiltonian (\ref{H}) becomes:
\be
H_{\rm SU(8)}\to
\sum_{\bk,\sigma\mu} \epsilon_{\bk}^0 \Psi_{\bk,\sigma}^{(\mu) \dagger}
\Psi_{\bk,\sigma}^{(\mu)}+ H_{int}'~, \label {Hsu8} 
\ee
where $\Psi^{(\mu) \dagger} = (e_1^\dag,e_2^\dag,h_1^\dag,h_2^\dag)$.
Ignoring the effects of bands away from the FS, the kinetic
part of 
$H_{\rm SU(8)}$ (\ref{Hsu8}) has an exact
SU(8) symmetry involving orbital ({\em both} electron
and hole) and spin degrees of freedom, 
$\mu$ and $\sigma$, respectively. This symmetry can be used 
to classify various vertices in $H_{int}'$ -- 
ultimately generated by $H_{int}$ (\ref{Hint}) which
itself has only U(1)$_{\rm charge}\times$SU(2)$_{\rm spin}$ symmetry -- and analyze 
various symmetry-breaking patterns, 
starting with $SU(8)\to SU(4)\times SU(4)$, as discussed in \cite{cvetkovicsu8}.

To illustrate the physics, we focus first on the {\em minimal} model:
SU(8) $\to $ SU(2). This leaves one with only two fermion flavors, 
$e$ and $h$. Note that {\em spin} SU(2) symmetry is suppressed, but
the {\em orbital} electron-hole symmetry remains, as it is essential
for this problem. We now obtain:
\be
H_{\rm SU(2)} &=&
\sum_{\bk} \epsilon_{\bk} [e_{\bk}^{\dagger} e_{\bk} +
h_{\bk}^{\dagger} h_{\bk}] +  \thalf \int \rmd^2\br  \rmd^2\br' \times  \nonumber \\
&&[ U^{e}(\br -\br')n^e(\br)n^e(\br') +
U^h(\br -\br')n^h(\br)n^h(\br') \nonumber \\
&&- 2W(\br -\br')n^e(\br)n^h(\br')] + H_{\rm SU(2)}^{\rm mixed}~, \label {Hsu2} 
\ee
where
$U^{e/h}(\br)$ and $W(\br)$ are the Fourier transforms of 
$\tilde V_{\bk - \bk'}\langle\zeta^{(e/h)}_{\bk,\bk'}\zeta^{(e/h)}_{\bk',\bk}\rangle_{\rm FS}$ and
$\tilde V_{\bk - \bk'}\langle\zeta^{(e)}_{\bk,\bk'}\zeta^{(h)}(\bk',\bk)\rangle_{\rm FS}$, respectively.
$\langle\cdots\rangle_{\rm FS}$ indicates the $\bk$- and
$\bk'$-dependence was replaced by the average over the FS -- this is justified later.
The SU(2) symmetry implies $U^{e}(\br)= U^{h}(\br)$.
Finally, the general $H^{\rm mixed}$ contains smaller mixed vertices and
eventually plays a prominent role in our theory; however,
we initially --- but only temporarily --- 
set it to zero,
$H_{\rm SU(2)}^{\rm mixed}\to 0$. 

%figure

%\onecolumn

%============= FIGURE BEGINS ===============

The intraband scattering $W (\br)$ has a minus sign in front of it.
This is the result of the p-h transformation and indicates
that, us having started with a (screened) Coulomb repulsion between
the original electrons, the $e$ and $h$ flavors now mutually {\em attract}.
Consequently, at low energies, 
$H_{\rm SU(2)}$ (without $H_{\rm SU(2)}^{\rm mixed}$)
is {\em equivalent} to 
the {\em negative} $U$ Hubbard model at low (or high) $x$, assuming that $W(\br)$ is 
short-ranged and $\epsilon_{\bk}$ can be represented
in the effective mass approximation 
($W(\br)\to W\delta (\br)$ and $\epsilon_{\bk}\to \bk ^2/2m$).
Both assumptions should be valid since $k_F\ll M$.
The ground state is
a ``superconductor'', with an anomalous correlator
$\langle e^\dag h^\dag\rangle\not = 0$, 
where $e^\dag (h^\dag )$ creates a 
``spin up (down)'' fermion $f_\uparrow$ ($f_\downarrow$).
For $T>T_c \sim E_F\exp (-1/N(0)W)$, the system is in its
normal state (see phase diagram in Fig. \ref{fig1}). 
At $T<T_c$ one enters a broken symmetry state, with 
``off-diagonal'' long range order and gapped fermions.

Of course, the above ``superconductivity'' is useful but entirely fictitious mathematical
construct, resulting from the p-h transformation used to enhance the
symmetry of our basic model for iron pnictides. Still, what is this 
``superconductivity'' in the real world?
By retracing our steps and undoing the p-h transformation,
the ``off-diagonal'' order in 
$\langle f^\dag_{\bk\uparrow} f^\dag_{-\bk\downarrow}\rangle =
\langle e^\dag_{\bk} h^\dag_{-\bk}\rangle$
translates into the diagonal density-wave, 
$\langle d^\dag_{\bk + {\bf M}} c_{\bk}\rangle\not = 0$,
connecting electrons from two pockets of the FS separated by
${\bf M}=(\pi,\pi)$; a 
{\em valley} density-wave (VDW). Note that the VDW describes
{\em both} spin  {\em and} charge/orbital density-waves (SDW/CDW/ODW). 
With the spin SU(2) symmetry suppressed in our minimal model, one 
cannot -- and should not -- distinguish
between the two. We identify the above VDW (SDW/CDW/ODW combination) formation
as the physical mechanism driving the 
density-wave orderings observed in numerous experiments.

We can pursue this VDW-``superconductor'' analogy a bit further:
in real FeAs materials, the electron-hole pockets are not identical, the main
difference being their distinct $k_F$'s. In our fictitious
superconductor, this translates to different ``chemical potentials'', 
$\mu^e\not = \mu^h$ for the electron
and the hole valleys. This is nothing but the external
Zeeman splitting in a fictitious negative $U$ Hubbard model. As
$\delta\mu= \mu ^e - \mu^h$ grows, the ``superconducting'' state approaches
the ``Chandrasekhar-Clogston'' limit, giving way
to a {\em non-uniform} Fulde-Ferrell-Larkin-Ovchinikov (FFLO) ground state
at an {\em incommensurate} wavevector ${\bf q}$, where
$|{\bf q}|$ is set by 
$\delta k_F=k_F^e-k_F^h$. This ``FFLO state''
is just an incommensurate (IC) VDW at the wavevector ${\bf M} + {\bf q}$.
Finally, as $\delta k_F~(x)$ exceeds certain critical value $\delta k_c~(x_c)$,
the ``superconducting'' state is destroyed and so is 
the VDW (SDW/CDW) in a real FeAs system (Fig. \ref{fig1}) \cite{latticeeffects}.

\section{Valley density-wave and interband superconductivity}

The above ``superconductor'' analysis dealt with an idealized model but
its main conclusions apply to the real Fe-pnictides:
i) the dominant instability is the VDW at wavevector ${\bf M}$, 
a unified spatially-modulated state manifested through the
{\em combined} SDW/CDW/ODW and 
a structural transformation \cite {OakRidge}, the details of which depend on
non-universal features of individual materials; 
ii) since hole and electron valleys are not identical, the VDW is the
p-h analog of a fictitious FFLO state, resulting 
almost always in portions of the FS which are {\em not gapped} (Fig. \ref{fig1}).
Consequently, the SDW/CDW/ODW's in FeAs are highly itinerant and
coexist with finite density of normal charge carriers, exhibiting
metallic conductivity \cite {PrFeAsO}; and, finally, iii)
the Hamiltonian $H_{\rm SU(2)}$ \eqref{Hsu2} {\em without}
the mixed vertices (i.e., with $H_{\rm SU(2)}^{\rm mixed}\to 0$) contains
only {\em three} basic ground states: fictitious uniform and non-uniform FFLO ``superconductor''
(C and IC VDW) and the normal state (Fig. \ref{fig1}). 
Thus, if purely electronic interactions are to have
a prominent role in generating Fe-based superconductivity of the 
real world, this effect {\em must} arise
from $H_{\rm SU(2)}^{\rm mixed}$. This is an important result, and an added
benefit of our transforming the original problem into a fictitious superconductor
\cite{Trubowitz}.

With this last point in mind, we now restore
these mixed vertices to investigate the {\em real} superconductivity in our
``superconductor'' model. 
It is
beneficial at this stage to add extra two flavors to the elementary $SU(2)$ model
and demand an additional global $SU(2)$ isospin 
symmetry with respect to these new flavors -- 
this isospin degree of freedom is completely inert and can be thought of as either the real
spin or an additional orbital index. Its role is purely mathematical as it couches
the following analysis in the language most easily translated to the ultimate
realistic description of pnictides \cite{Unonlocal}.
Furthermore, viewing this isospin as simply the real spin is useful since,
assuming a reasonable degree of total spin conservation in pnictides, the additional
SU(2) global symmetry limits  the number of terms in $H^{\rm mixed}$ that need to be
considered. 
The mixed vertices allowed by this extension of our model are
mixed scattering $G_1$, and Josephson-type term $G_2$ which,
in absence of nonlocal interactions, has to be in a spin-singlet channel
\be
  H^{\rm mixed}_{\rm spin} &\sim& G_1 c_{\sigma}^\dagger d_{\sigma}
    d_{\sigma'}^\dagger c_{\sigma} + \nonumber \\
    && \thalf G_2 (\sigma c_{\sigma}^\dagger c_{-\sigma}^\dagger)
    (\sigma' d_{-\sigma'} d_{\sigma'} ) + h.c.~ . \label {H_SU2_mixed}
\ee
The relation of these vertices to the screened Coulomb interaction 
is given in Eq.\ \eqref {G1}.
We assume that the corresponding coupling constants are in the regime
$G_1, G_2\ll W, U_h, U_e$, as will be justified momentarily.

These preliminaries in place, we are ready to answer the key question:
what is the effect of finite $G_1, G_2$ on the previous analysis?
Analyzing corrections in the perturbation theory to
our four point vertices, we find contributions to the processes which
are dependent on whether the incoming, and outgoing, spins 
are parallel or not. This occurs due to the fact that incoming or outgoing
states of mixed term $G_2$ as well as the intraband interaction $U$,
are spin singlets of either holes or electrons,  Eq.\ \eqref {H_SU2_mixed}.
The interband scattering term is therefore conveniently split into two pieces
\be
  W d_{\sigma'}^\dagger c_{\sigma}^\dagger c_{\sigma} d_{\sigma'}
    \longrightarrow W' d_{-\sigma}^\dagger c_{\sigma}^\dagger c_{\sigma} d_{-\sigma} +
    W'' d_{\sigma}^\dagger c_{\sigma}^\dagger c_{\sigma} d_{\sigma},
\ee
with the bare values for both coupling being identical to $W$. The first mixed term
$G_1$ is split in identical fashion, while the intraband scattering and the
$G_2$ mixed term are required to scatter particles with incoming opposite spins
and therefore do not need to undergo the same separation (or equivalently, 
$G_2'' \equiv 0$, etc.).
At the lowest order, we find that different types
of vertices receive the following corrections in the perturbation theory:
\be
  g_U(\omega) &=& g_U - g_U^2\ln \bigl(\frac{\Lambda}{\omega}\bigr)_{pp} -
    g_2^2\ln \bigl(\frac{\Lambda}{\omega}\bigr)_{pp},  \nonumber \\
  g_2(\omega) &=& g_2 - 2 g_2 g_U \ln \bigl(\frac{\Lambda}{\omega}\bigr)_{pp} +
    2 g_2 g'_W \ln \bigl(\frac{\Lambda}{\omega}\bigr)_{ph}^c + \nonumber \\
    && 2 g_2 g''_W \ln \bigl(\frac{\Lambda}{\omega}\bigr)_{ph}^v
    - 2 g_2 g''_1\ln \bigl(\frac{\Lambda}{\omega}\bigr)_{ph},  \nonumber \\
  g'_W(\omega) &=& g'_W + (g'_W)^2 \ln \bigl(\frac{\Lambda}{\omega}\bigr)_{ph} +
    g_2^2\ln \bigl(\frac{\Lambda}{\omega}\bigr)_{ph},  \nonumber \\
  g''_W(\omega) &=& g''_W + (g''_W)^2 \ln \bigl(\frac{\Lambda}{\omega}\bigr)_{ph}, \nonumber \\    
  g'_1(\omega) &=& g'_1 - 2 g'_1 g''_1 \ln \bigl(\frac{\Lambda}{\omega}\bigr)_{ph} 
    + 2 g'_1 g''_W ln \bigl(\frac{\Lambda}{\omega}\bigr)_{ph}^v, \nonumber \\
  g''_1(\omega) &=& g''_1 - (g'_1)^2 \ln \bigl(\frac{\Lambda}{\omega}\bigr)_{ph} -
    (g''_1)^2 \ln \bigl(\frac{\Lambda}{\omega}\bigr)_{ph} - \nonumber \\
    &&g_2^2\ln \bigl(\frac{\Lambda}{\omega}\bigr)_{ph} + 2 g''_1 g''_W \ln \bigl(\frac{\Lambda}{\omega}\bigr)_{ph}^v,
\label{RG}
\ee
where $g_U$,  $g_2$, $g'_W$, $g''_W$, $g'_1$ and $g''_1$ are just the 
vertices $U (=U_h=U_e)$ \cite{Unonlocal}, $G_2$, $W'$, $W''$, $G'_1$ and $G''_1$,
respectively, measured in units of inverse density-of-states (DOS) at the
Fermi level. The logarithmic divergences in (\ref{RG}) arise  from two sources:
first, the standard Cooper pairing instability in the particle-particle (pp) channel and, second, 
the perfect nesting
of the hole and electron bands in the particle-hole (ph) channel, i.e.,
our fictitious  ``Cooper'' instability. Finally, 
$\ln \bigl(\frac{\Lambda}{\omega}\bigr)_{ph}^{c(v)}$ denotes a 
crossing (vertex) diagram in the p-h channel --- it strictly diverges
only in the $k_F^{e,h}/M\to 0$ limit and is otherwise finite.
Needless to say, there are many additional terms that contribute to various vertices
at the leading order in perturbation theory. However, all such terms are finite
in the low energy limit and are omitted from (\ref{RG}).

The third and fourth lines of \eqref{RG} are 
just the mathematical shorthand for our earlier discussion:
under renormalization, the coupling constants $g'_W$ and $g''_W$ keep growing,  ultimately 
generating the VDW instability, driven by $W$ (the short-ranged attraction of our fictitious
Hubbard model). We notice, however, that $G_2$ enhances
the growth of $W'$, thereby giving slight edge to SDW (spin-triplet)
over a CDW (spin-singlet). The first and last two lines tell us 
that $U$ and $G_1$'s do not interfere:
the intraband repulsion $g_U$, initially large, is rapidly 
renormalized downwards, toward the Fermi liquid behavior. 
$g_1$'s are typically small to begin
with and are also driven down;
in practice, they can be set to zero. The interesting physics is reserved
for $G_2$. The growth of $g_W$'s fuels the growth of $g_2$ and thus the mixed vertex
describing the resonant pair scattering between the hole and electron bands --- i.e., 
the ``Josephson'' interband vertex $c^\dag d c^\dag d + {\rm h.c.}$ ---
becomes strongly enhanced as one approaches the VDW (SDW/CDW/ODW) instability. However,
since typically $G_2\ll W$ (see below), $g_W$ wins, resulting in the VDW order.
Once the VDW is formed, the fermions are gapped and the singular behavior disappears, and
with it any additional enhancement of $G_2$.

The situation changes, however, when the differences in size between $h$ and $e$ bands
are included, i.e., when the ``Zeeman splitting'' is turned on,
by doping or pressure in real FeAs (Fig.  \ref{fig1}).
This cuts off the fictitious ``Cooper'' divergence, resulting in our ``FFLO'' state
and eventual disappearance of VDW. In this case, the portions or even all of the FS is
still available for the true Cooper pairing and the real superconductivity becomes
a viable option. 
The remarkable feature of interband pair resonance is that it can produce 
real superconductivity {\em irrespective} of its sign \cite {Suhl}. Thus, strongly
enhanced $G_2$ can take advantage of the real-world Cooper singularity -- which is
always present -- and amplify a preexisting {\em intraband} superconducting 
instability or generate one entirely on its own. We will revisit this point shortly.

\section{Valley density-wave and superconductivity in real iron pnictides}

This is as far as we can go within the idealized picture: we now must face up to
the complexities of the real materials. First,
there are four, two $h$ and two $e$, bands which deviate from an
ideal parabolic shape and
whose $k_F$'s are different and, second, all vertices -- intra, interband
and mixed -- have considerable structure as one moves over different portions
of the FS. The latter is an important point and reflects a fundamental feature
of FeAs: all five d-orbitals need to be included in realistic calculations and 
various two- and three-orbitals models will fall short in addressing the
phenomenology of real materials. We find that, no more than a single orbital,
$d_{2z^2-x^2-y^2}$, can be dropped without a major qualitative
disruption of the character of the electronic
states at the FS; thus a four-orbital description is the absolute minimum.
Finally, the lattice effects produce modifications
to our continuum picture which need to be addressed \cite{latticeeffects}.

\begin{figure}[tbh]
\centering
\raisebox {0.7in} {a)}
\includegraphics[width=0.3\columnwidth]{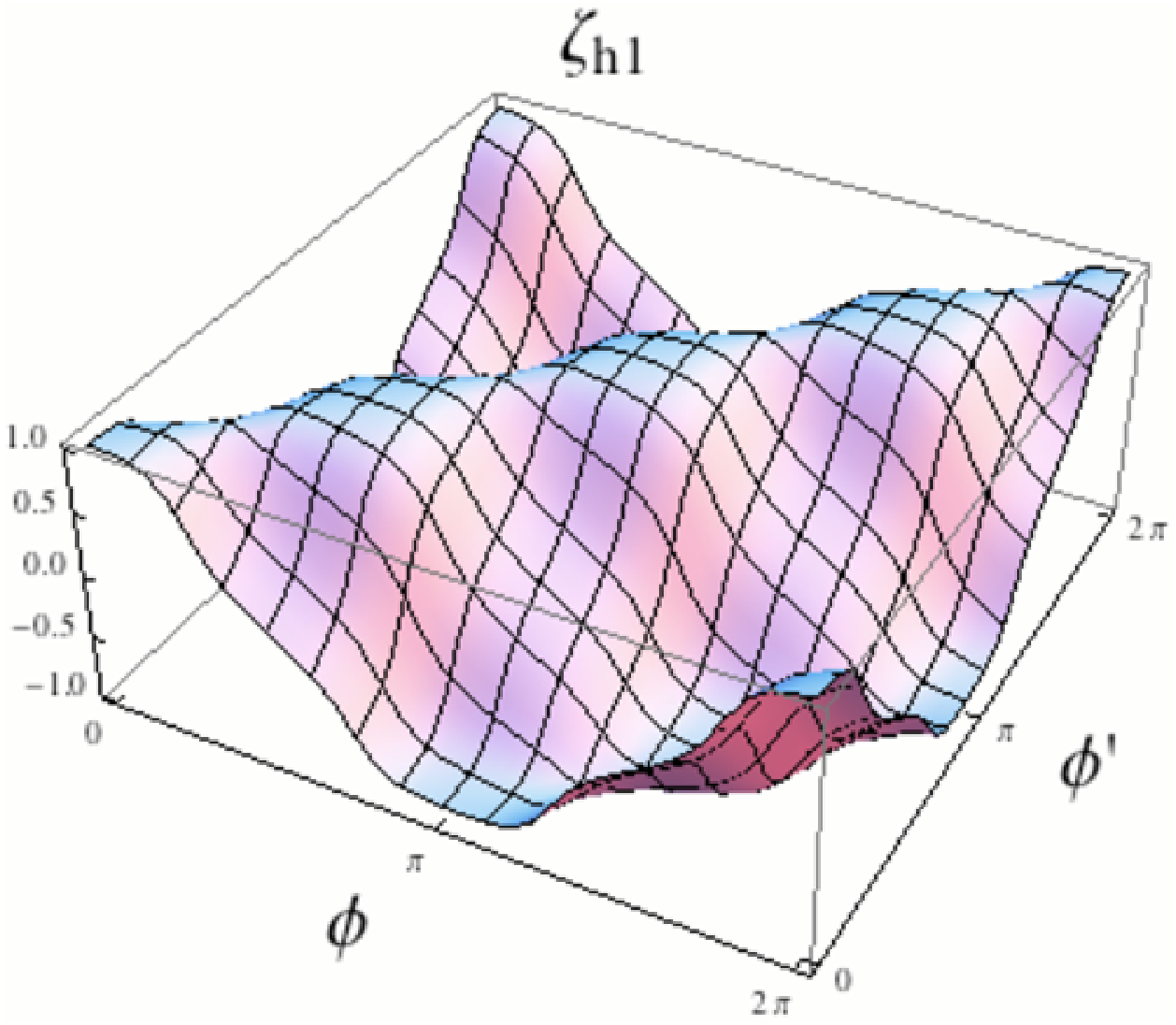}
\includegraphics[width=0.3\columnwidth]{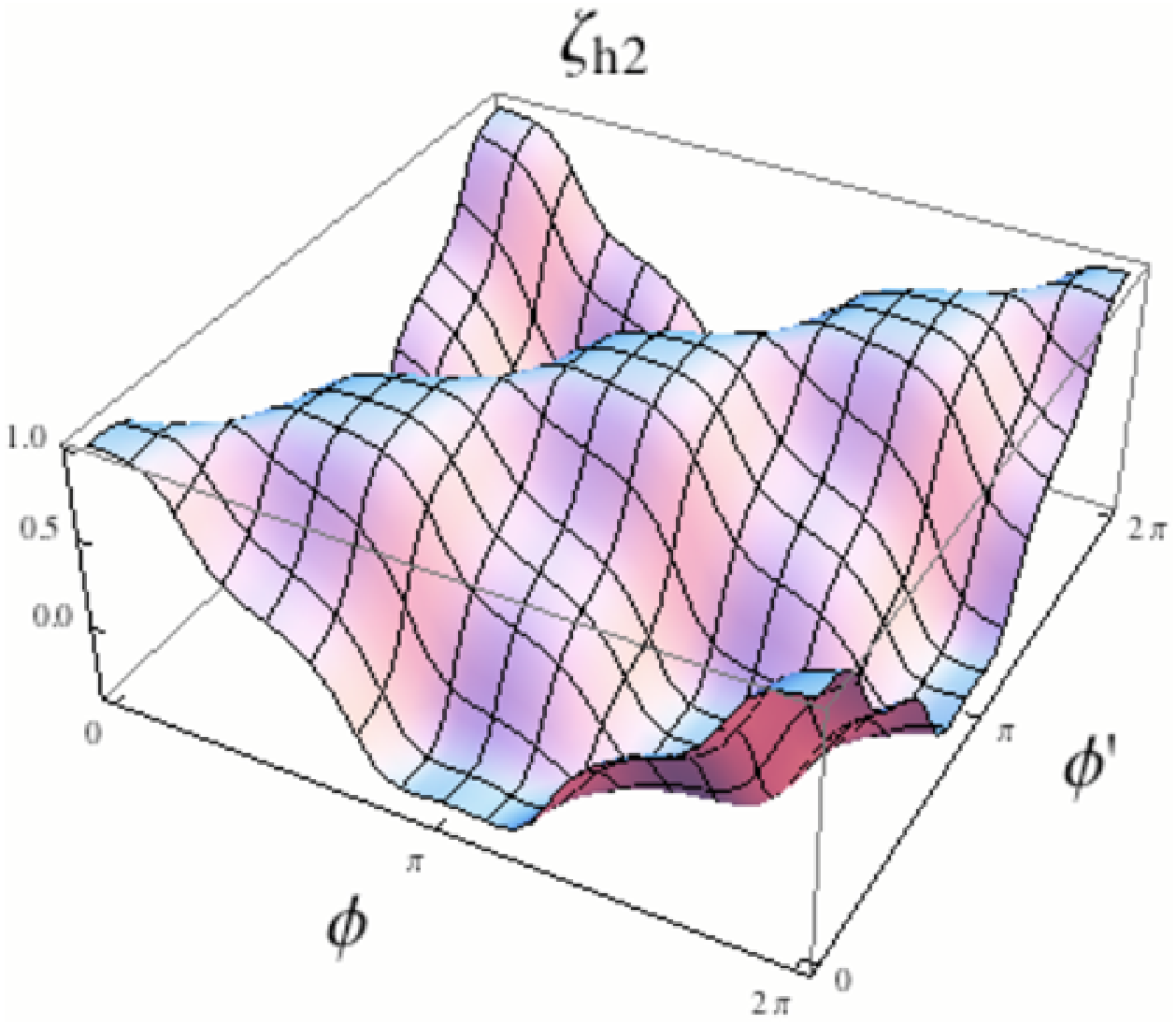}
\includegraphics[width=0.3\columnwidth]{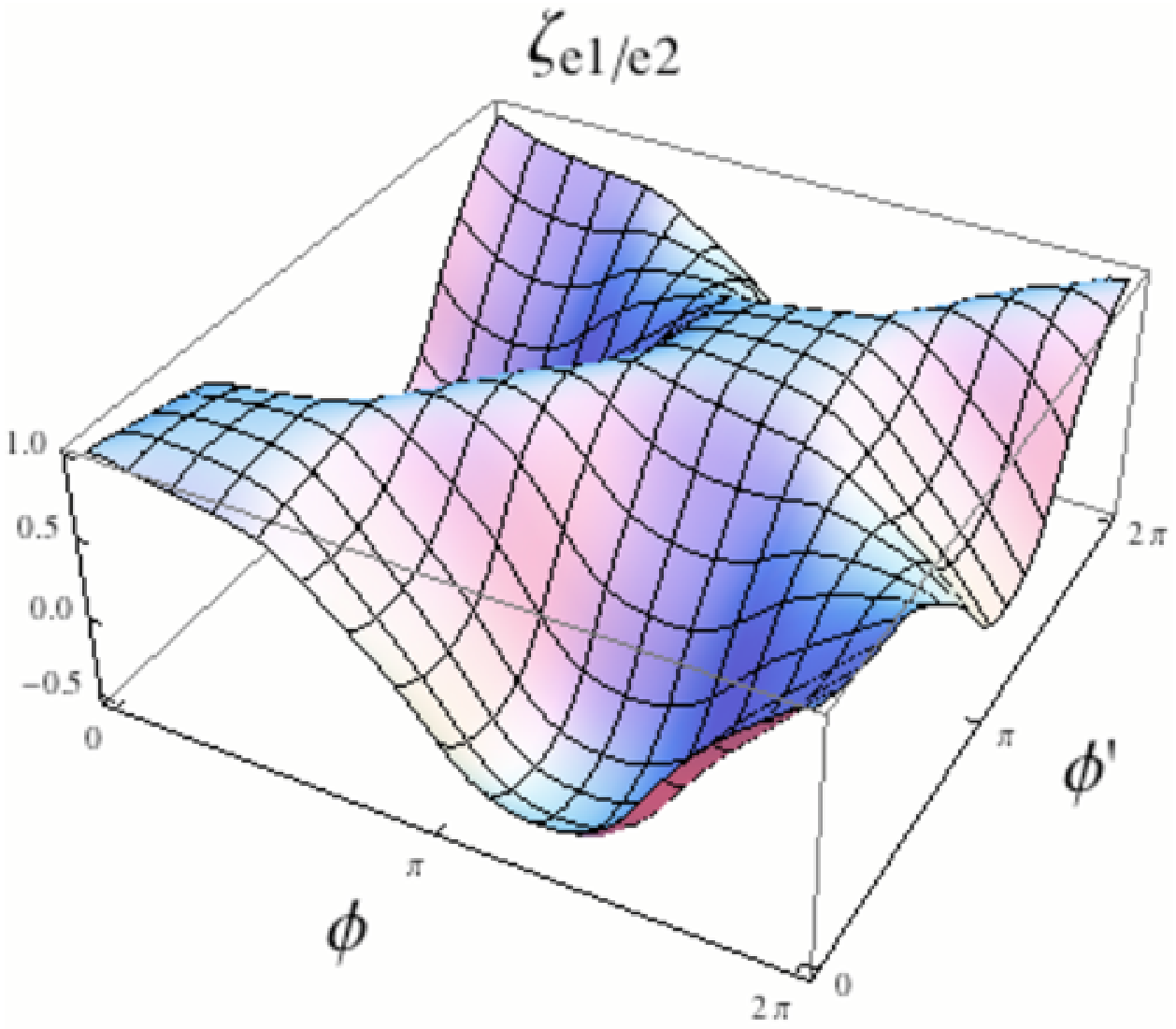}
\\
\raisebox {0.7in} {b)}
\includegraphics[width=0.3\columnwidth]{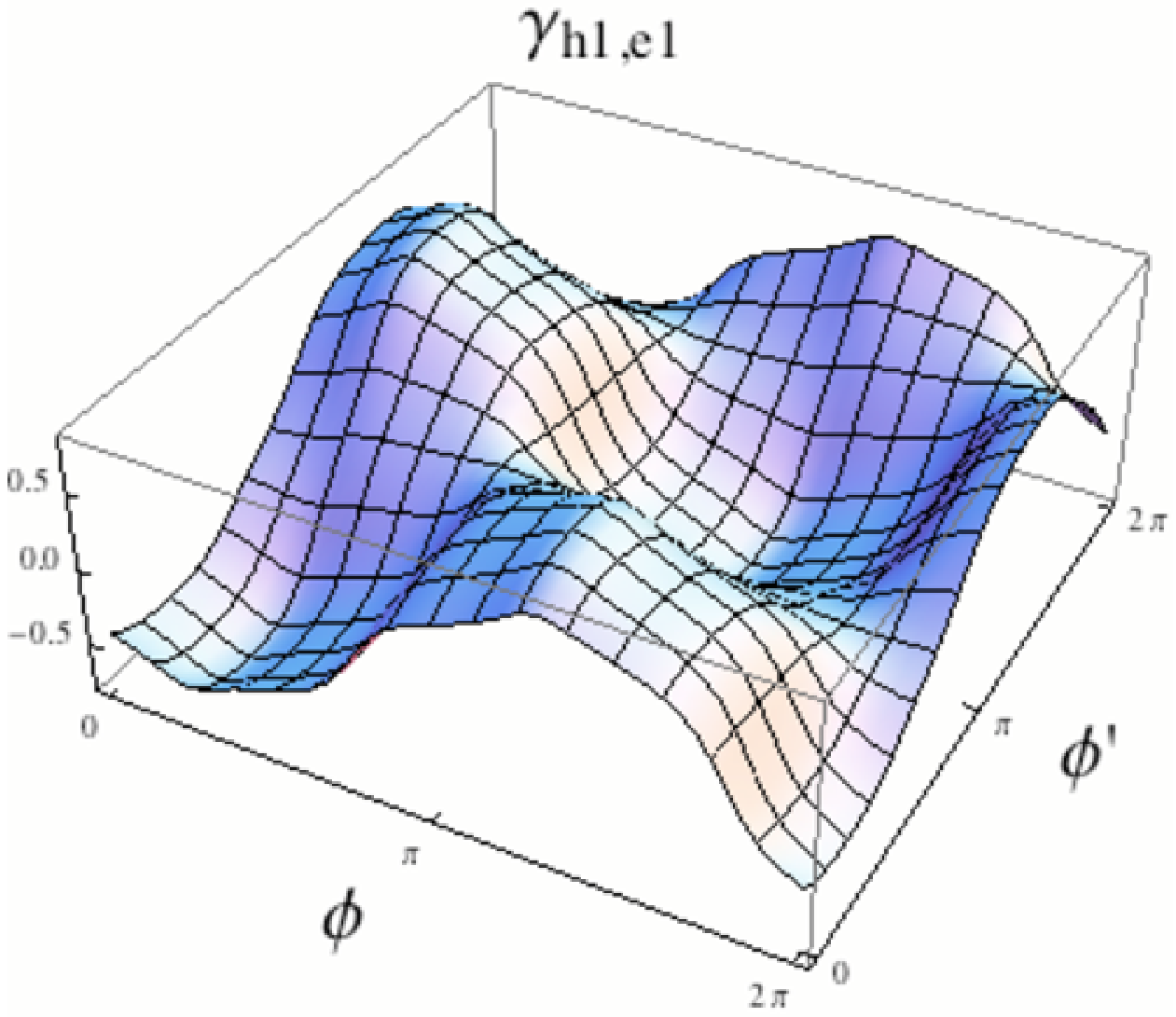}
\includegraphics[width=0.3\columnwidth]{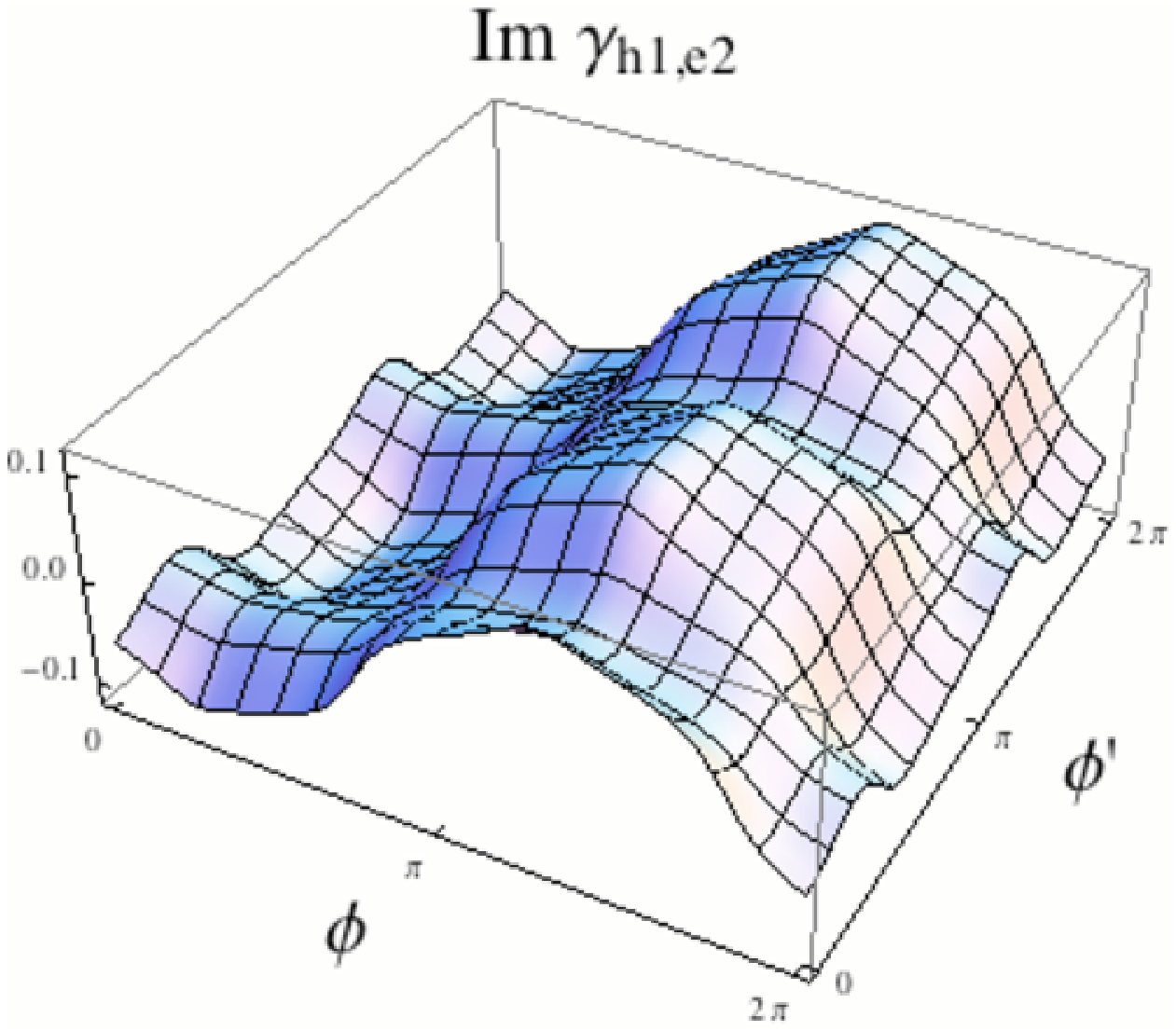}
\caption{\label{FigZetas} (Color online)
a) inter/intraband $\zeta^{(e / h)}_{\bf k, \bf k'}$
of Eq. \eqref{density} (in $k_F\ll M$ limit) 
for the inner and outer hole
bands, and the electron bands as $\bk$
and $\bk'$ go around respective FSs. 
b) mixed $\gamma^{(e, h)}_{\bk, \bk'}$ \eqref{density} are clearly smaller than $\zeta$'s.
}
\end{figure}

We use the full 8+8-band tight-binding model \cite {Cvetkovic} to
find the electron ($\varphi_{\bk}^{(\alpha)}$) and 
hole ($\phi_{\bk}^{(\beta)}$) wavefunctions.
This model yields $\zeta$'s and $\gamma$'s \eqref{density} 
shown in Fig.\ \ref {FigZetas}. For a fixed
$\bk$, a given $\zeta$ varies as $\bk'$ goes around the FS. At $\bk = \bk'$,
the normalization of wave-functions sets $\zeta$ to 1. As $\bk$ and $\bk'$
move apart, so does $\zeta$ decreases until it reaches its minimum
at $\bk' = - \bk$. Based on the symmetry properties
of the atomic orbitals, one obtains ($k_F\ll M$)
\be
  \zeta^{(\alpha)}_{\bk, -\bk} = \pm (\sum_{\mu \in even} | b^{(\alpha)}_{\mu} |^2 -
  \sum_{\mu \in odd} |b^{(\alpha)}_\mu|^2), \label {zeta_opposite}
\ee
where $b_\mu^{(\alpha)}$ is the amplitude
of atomic orbital $\mu$ in a hole state $(\alpha)$ or, 
equivalently, an electron state $(\beta )$.
Each orbital's contribution is determined by its in-plane parity 
(i.e., sign change under $(x, y, z) \to (-x, -y, z)$); 
even/odd orbitals contribute with $+/-$, 
or vice versa. Our model uses orbitals of different parity, and
consequently, \eqref {zeta_opposite} is bound between
$-1$ and 1, the precise value depending on the amount of mixing of even and 
odd orbitals within a state. For example, compare $\zeta^{(h1)}$ and $\zeta^{(h2)}$
(Fig.\ \ref {FigZetas}). Since both hole bands have a significant contribution of
$d_{xz/yz}$ atomic orbitals (odd), both are similarly shaped. The 4-fold repetitive structure
is due to the $C_4$ lattice symmetry. However, the minimum values ($\bk = -\bk'$) are
different: $\zeta^{(h1)}$ nearly reaches $-1$, whereas $\zeta^{(h2)} \gtrsim -0.6$.
This reflects the fact that the outer
hole band possess a significant overlap with $d_{xy}$ (even) orbital state,
while the inner hole band is almost entirely made of odd bands.
In a more limited model, where only
bands of a certain parity are kept \cite {Raghu},
a topological `Berry phase winding' can be defined for each section
of the FS \cite {Berkeley1}.
Depending on this `winding', $\zeta_{\bk, -\bk}$ would have
to be either +1 or $-1$, and the consequences of the latter would include a suppression of
an $s$-wave VDW in a favor of a $p$-wave one. Within our model, this
notion of topology is absent.

%============= FIGURE BEGINS ===============

Next, the above form factors $\zeta$'s and $\gamma$'s
are used to compute all the interaction vertices 
(intra, interband and mixed) stemming from \eqref{Hint} along different
sections of the FS ($h1$, $h2$, $e1$, and $e2$) and to extract 
the corresponding coupling constants in the $C_4$ ``angular momentum'' channels,
$s$, $p$, and $d$. The results, normalized by the
overall strength of the screened Coulomb interaction in \eqref{Hint},  
are
\be
\begin {tabular} {c | c c c c | c c c c}
  & \multicolumn{4}{|c}{$U$} & \multicolumn {4}{|c}{$W$} \\
  & \tiny (h1) & \tiny (h2) & \tiny (e1) & \tiny (e2) & \tiny (h1,e1) & \tiny (h1,e2) &
  \tiny (h2,e1) & \tiny (h2,e2) \\ 
  \hline
  $s$ &
  \footnotesize 0.44 & \footnotesize 0.31 & \footnotesize 0.36 & \footnotesize 0.35 &
  \footnotesize 0.21 & \footnotesize 0.25 & \footnotesize 0.27 & \footnotesize 0.29 \\
  $p_x$ &
  \footnotesize 0.02 & \footnotesize 0.10 & \footnotesize 0.08 & \footnotesize 0.10 &
  \footnotesize 0.11 & \footnotesize 0.10 & \footnotesize 0.11 & \footnotesize 0.11 \\
  $p_y$ &
  \footnotesize 0.02 & \footnotesize 0.10 & \footnotesize 0.09 & \footnotesize 0.10 &
  \footnotesize 0.11 & \footnotesize 0.10 & \footnotesize 0.11 & \footnotesize 0.11 \\
  $d_{x^2-y^2}$ &
  \footnotesize 0.14 & \footnotesize 0.08 & \footnotesize 0.02 & \footnotesize 0.03 &
  \footnotesize 0.06 & \footnotesize 0.07 & \footnotesize 0.05 & \footnotesize 0.06 \\
  $d_{2xy}$ &
  \footnotesize 0.08 & \footnotesize 0.04 & \footnotesize 0.07 & \footnotesize 0.07 &
  \footnotesize 0.05 & \footnotesize 0.06 & \footnotesize 0.04 & \footnotesize 0.05
\label{tableU}
\end {tabular}
\ee
\be
\begin {tabular} {c | c c c c | c c c c}
  & \multicolumn{4}{|c}{$G_1$} & \multicolumn {4}{|c}{$G_2$} \\
  & \tiny (h1,e1) & \tiny (h1,e2) & \tiny (h2,e1) & \tiny (h2,e2) &
   \tiny (h1,e1) & \tiny (h1,e2) & \tiny (h2,e1) & \tiny (h2,e2) \\ 
  \hline
  $s$ &
  \footnotesize 0.11 & \footnotesize 0.00 & \footnotesize 0.00 & \footnotesize 0.00 &
  \footnotesize 0.15 & \footnotesize 0.01 & \footnotesize 0.11 & \footnotesize 0.02 \\
  $p_x$ &
  \footnotesize 0.00 & \footnotesize 0.00 & \footnotesize 0.00 & \footnotesize 0.00 &
  \footnotesize 0.00 & \footnotesize 0.00 & \footnotesize 0.01 & \footnotesize 0.00 \\
  $p_y$ &
  \footnotesize 0.00 & \footnotesize 0.00 & \footnotesize 0.00 & \footnotesize 0.00 &
  \footnotesize 0.00 & \footnotesize 0.00 & \footnotesize 0.01 & \footnotesize 0.00 \\
  $d_{x^2-y^2}$ &
  \footnotesize 0.00 & \footnotesize 0.00 & \footnotesize 0.04 & \footnotesize 0.00 &
  \footnotesize 0.00 & \footnotesize 0.00 & \footnotesize 0.00 & \footnotesize 0.00 \\
  $d_{2xy}$ &
  \footnotesize 0.05 & \footnotesize 0.00 & \footnotesize 0.00 & \footnotesize 0.00 &
  \footnotesize 0.03 & \footnotesize 0.00 & \footnotesize 0.00 & \footnotesize 0.00
\label{tableG}
\end {tabular}
\ee
All the vertices \eqref{tableU} and \eqref {tableG} are given
in the original $c,d$ electron basis (\ref{H}-\ref {Hint}) and are all positive (repulsive);
they are easily converted into the basis used in \eqref{Hsu8} by the p-h
transformation (i.e., $W\to -W$, etc.).
The numbers in the above table change if additional forms of interaction in real
space are considered, for example those of Eq. \eqref{HintWannier}, as discussed earlier. 
Again, provided one is outside the regime dominated by the Hund's coupling, such changes
are minor. 

Obviously, when it comes to the interband vertex ($W$) as well as
all the other vertices,
the ``s-wave'' channel dominates, which retroactively justifies
our earlier idealized analysis in terms of an attractive Hubbard model
and a fictitious ``superconductor.'' We find that -- depending on the
overall strength of Coulomb repulsion --
the most likely ground state is a multiband 
VDW (SDW/CDW/ODW) which, due to the mismatch of the hole and electron bands and
the underlying lattice effects, is generically in the ``FFLO'' region of the
phase digram in Fig. \ref{fig1}, leaving portions of the original FS ungapped
and metallic. As expected, this VDW (SDW/CDW/ODW) symmetry breaking at 
wavevector ${\bf M}$ is fueled by a large susceptibility of nearly-nested hole and electron 
valleys \cite {Cvetkovic}.

If the Coulomb repulsion is just below what is needed to produce a metallic VDW 
(Fig. \ref{fig1}), the mixed ``Josephson'' vertex $G_2$ is strongly enhanced,
as illustrated by \eqref{RG} and surrounding discussion. This is the regime
where the interband superconductivity \cite{Mazin,Cvetkovic,Chubukov} is possible. Here, 
an important point needs to be made: there are {\em two} ways in which $G_2$
can lead to high temperature superconductivity in Fe-pnictides: first,
$G_2$ {\em itself} can be {\em the source} of superconductivity. 
This is a naturally appealing
theoretical scenario, since it relies on the proximity to a VDW (SDW) 
instability to enhance $G_2$ 
and uses purely electronic interactions to
generate superconducting order. The difficulty in this case is that $G_2$ has to overwhelm 
the {\em intraband} Coulomb repulsion $U^{e}$ and $U^{h}$ 
before superconductivity becomes possible, the condition being 
rougly $G_2 > \sqrt{U^{e}U^{h}}$.
While $G_2$ takes off as one approaches the VDW (SDW/CDW/ODW), 
there is a reflection of this enhancement
in the renormalized values of $U^{e}$ and $U^{h}$ as well. For the realistic 
model with four ineqivalent bands
and the interaction vertices displayed in Tables \eqref{tableU} and \eqref {tableG}, 
this balancing act between $U$'s and $G_2$ 
becomes very sensitive, particularly since the bare $U$'s start as generically
larger and only two of $G_2$'s are not negligible in size while all 
four $U$'s are appreciable. Any effort to extend Eqs. \eqref{RG} to four realistic
bands and to all (intra, interband and mixed) vertices quickly descends into impenetrable
numerics with the above sensitivity to the bare values in  (\ref{tableU} - \ref {tableG}),
making it difficult to reach firm quantitative conclusions. A notable recent
progress along these lines was made in Refs.\ \cite{berkeleyRG}.

However, there is a reasonably straightforward way to illustrate 
the {\em qualitative} argument
for the interband superconductivity mechanism near the VDW phase boundary. This argument
follows straight from Eq. \eqref{RG}. Imagine that our isospin label is simply an
ordinary spin. Therefore, we have spinful electrons with two (instead
of four in real pnictides) orbital flavors, $c$ and $d$ ($h$ and $e$). In this case, $G_2$ is the 
interband pairing resonance in the spin-singlet channel. As argued earlier, we can safely
set $G_1=0$ and rewrite the remaining parts of \eqref{RG} as
\be
  \dot g_U &=& -g_U^2 - g_2^2,  \nonumber \\
  \dot g'_W &=& (g'_W)^2 + g_2^2, \nonumber \\
  \dot g''_W &=& (g''_W)^2, \nonumber \\
  \dot g_2 &=& - 2 g_2 g_U  + 2 g_2 (g'_W + g''_W) ~,
\label{RG'}
\ee
where $g$'s are functions of $\ln\bigl(\frac{\Lambda}{\omega}\bigr)$ and
$\dot g\equiv dg/d\ln\bigl(\frac{\Lambda}{\omega}\bigr)$ and we again
assume $U_h=U_e=U$.

Imagine for the moment that there is no last term in the last line of \eqref{RG'},
i.e., $W$'s and $G_2$ are not directly coupled. As one moves to low energies
$\omega\to 0$, $g_W$'s rapidly
grow and one ultimately reaches the point where the system turns into a VDW.
Meanwhile, $U$ and $G_2$ do nothing: this is easily seen by adding and subtracting 
the first and the last 
lines of \eqref{RG'} $\dot g_U \pm \dot g_2= -\bigl(g_U \pm g_2\bigr)^2$.
This is just the lowest order description of the real Cooper pairing instability 
in the s-wave spin-singlet channel, with the superconducting
gap parameter having either the same
sign for both bands $c$ and $d$ (conventional s-wave) or the opposite sign
(s$_\pm$-wave or an extended s-wave, s'), the corresponding coupling constants being
$g_U +g_2$ and $g_U-g_2$, respectively. Both of these coupling constants
are repulsive and thus both scale toward zero and into the Fermi liquid regime, leaving
the VDW and $W$ to determine the physics at low energies, {\em unless} 
$G_2 > U (\sim\sqrt{U_hU_e})$ at the {\em bare} level. This is not impossible, but
appears to be unlikely within the regime of interactions considered here, where
$G_2$ is typically quite a bit smaller than $U$. This tells us that the direct
coupling of $G_2$ to $W$'s in the last line of \eqref{RG'} must be {\em crucial}: the growth of
$W$'s as we approach the VDW eventually pulls $G_2$ along with it, 
while $U$ still continues being renormalized downward. This growth 
of $G_2$ generated by its coupling
to $W$'s and the VDW could ultimately result in
$G_2^\star > U^\star (\sim\sqrt{U_h^\star U_e^\star })$,
where $G_2^\star$ and $U^\star$ are the renormalized coupling constants at 
some low energy scale $\omega_0 \ll \Lambda$, even though 
$G_2 < U (\sim\sqrt{U_hU_e})$ at the bare level. This implies that the
coupling constant $g_U-g_2$ is {\em attractive} for $\omega <\omega_0$
and translates into growth of s$_\pm$ (s') pairing correlations at yet lower energies.
All of this is for nothing, however; $W$'s and the VDW instabilities are still
far stronger. But, if the strong growth of $W$ and the VDW instability are cut off
by our fictitious ``Zeeman splitting'' in doped or pressurized  FeAs (Fig.  \ref{fig1}), then $W$ stops growing at some
energy scale $\omega_z$ directly tied to the difference  
in FS size between $h$ and $e$ bands
and the corresponding lack of perfect nesting. 
Then, if $\omega_z < \omega_0$, the subleading
instability would take over and the ground state would be an s$_\pm$ (s') superconductor,
either adjacent to the VDW boundary or coexisting with it in the pockets left ungapped
by the VDW (Fig.  \ref{fig1}). The above argument is similar in spirit to the weak
coupling mechanism for d-wave superconductivity once the single band repulsive Hubbard
model is doped away from half-filling and the SDW ground state. There is an
important difference, however: if $\omega_z > \omega_0$ there will be no superconducting
ground state since $g_U-g_2$ is still repulsive. This is a qualitative point and it
underscores the fact that an s$_\pm$ (s') superconductor still has an overall
s-wave symmetry and, unlike the nodal d-wave, must contend with the strength of the
bare intraband repulsion.

The second way is now obvious: $G_2$ enhancement near the VDW instability
can overcome the repulsion $U$ if an {\em attractive} intraband interaction
is at work as well. Such intraband attraction might 
come from phonons, for example. This attraction
may or may not suffice to produce superconductivity  by itself -- the key point is that
it reduces the effective $U$'s (\ref{tableU})
allowing the enhanced $G_2$ to cross over the hurdle. Note that in both of these
cases, the purely electronic and the phonon-assisted one, 
the superconducting gap on the hole and the electron portions of the FS will 
have the opposite sign \cite{Mazin,Cvetkovic,Chubukov}, reflecting the fact that the 
interband pairing term $G_2$ is repulsive.

\section{Conclusions}

In summary, we have proposed an idealized model of Fe-pnictides
which includes an electron and a hole band,
and takes advantage of their similar shape and size. The p-h
transformation maps this model into a fictitious attractive
Hubbard model in external ``Zeeman'' field. The ground states, fictitious
superconductor and the ``FFLO'' state, correspond to insulating and metallic VDW
in real materials. 
Next, by considering 
deviations from perfect nesting, two hole and two electron bands, and other 
realistic features of Fe-pnictides, we analyze the structure of interactions in
the 8+8 orbital model \cite{Cvetkovic} and identify the interband pair resonance mechanism 
that can generate the {\em real} superconductivity in the region of the
phase diagram of Fe-pnictides where the VDW
order gives way to strong VDW fluctuations.

We thank I.\ Mazin, A.\ V.\ Chubukov, M.\ Kulic, C.\ Broholm, and W.\ Bao 
for useful discussions. This work was supported in part 
by the NSF grant DMR-0531159. Work at the Johns Hopkins Institute 
for Quantum Matter was supported by the U. S. Department of 
Energy Office of Science under Contract No. DE-FG02-08ER46544.

%reference

\bibliographystyle{apsrev}

\begin{thebibliography}{10}
\bibitem {Unonlocal} The introduction of the extra isospin flavor
defines a minimal model within which
all the interaction vertices are allowed to assume
their local form. For example,  the short-range, on-site, intraband (intraflavor)
repulsion $U_{h(e)}$ does not enter within our elementary $SU(2)$
model alone since $n_i^2=n_i$ for fermion particle number operators. Of course, this is not
a problem at all and one simply has to look at the non-local intraband repulsion and pairing
terms for spinless fermions but this is not what will ultimately correspond to actual
pnictides, where there are two extra pairs of flavors, additional orbital and the real
spin, for the total of four, and the local interactions 
among these flavors is all that is needed.
\end{thebibliography}

\end {document}